\pdfoutput=1
\documentclass[twocolumn]{aastex631}
\graphicspath{{./}{figures/}}
\usepackage[T1]{fontenc}

 \usepackage{tabularx}
 \usepackage{booktabs}
 \usepackage{array}

\usepackage{soul}
\usepackage{amsmath}
\usepackage{rotating} 

\shorttitle{Multi-epoch $H\alpha$ Spectropolarimetry of in Herbig Ae/Be Stars}
\shortauthors{L.~K. Dewangan et al.}

\begin{document}
\title{Temporal Evolution of Inner Circumstellar Geometry in Herbig Ae/Be Stars Revealed by Multi-epoch $H\alpha$ Spectropolarimetry}

\author[0000-0001-6725-0483]{L.~K. Dewangan}
\affiliation{Astronomy \& Astrophysics Division, Physical Research Laboratory, Navrangpura, Ahmedabad 380009, India.}
\email{lokeshd@prl.res.in}

\author[0000-0001-7340-8873]{Arijit Maiti}
\affiliation{Astronomy \& Astrophysics Division, Physical Research Laboratory, Navrangpura, Ahmedabad 380009, India.}
\affiliation{Indian Institute of Technology Gandhinagar Palaj, Gandhinagar 382355, India.}

\author{Sube Singh Gurjar} 
\affiliation{Astronomy \& Astrophysics Division, Physical Research Laboratory, Navrangpura, Ahmedabad 380009, India.}
\affiliation{Indian Institute of Technology Gandhinagar Palaj, Gandhinagar 382355, India.}

\author{Mudit K. Srivastava} 
\affiliation{Astronomy \& Astrophysics Division, Physical Research Laboratory, Navrangpura, Ahmedabad 380009, India.}
%
\begin{abstract}
We present multi-epoch $H\alpha$ spectropolarimetric observations of three Herbig Ae/Be stars, HD~58647, MWC~120, and MWC~147, obtained at epochs separated by $\sim$1--1.3~yr. By combining $H\alpha$ line profiles with velocity-resolved linear polarization and $Q$--$U$ diagrams, we investigate temporal changes in their inner circumstellar environments. The $H\alpha$ profiles retain broadly similar morphologies and overall velocity extents between epochs, although noticeable variations in line strength and peak intensity are observed. In contrast, the polarization spectra and $Q$--$U$ trajectories exhibit pronounced changes in morphology, orientation, and velocity-dependent behavior. The three sources display distinct temporal characteristics: HD~58647 evolves from a compact, relatively weakly structured trajectory to a pronounced, predominantly linear excursion; MWC~120 exhibits open, structured trajectories at both epochs, with changes in their orientation and distribution; and MWC~147 evolves from a complex, vertically extended trajectory to a more confined and comparatively coherent, approximately linear arrangement. In HD~58647, strong velocity-dependent variations in polarization across the $H\alpha$ profile suggest a structured spatial relationship between the line-emitting and scattering regions. These source-dependent variations demonstrate that substantial changes in velocity-resolved polarization morphology can occur without corresponding changes in the overall $H\alpha$ velocity extent. This indicates that the projected polarization structure can evolve independently of the kinematic extent of the $H\alpha$-emitting region. Multi-epoch, velocity-resolved spectropolarimetry therefore provides a sensitive probe of temporal variations in the innermost circumstellar environments of intermediate-mass young stars, revealing variations that may not be apparent from spectroscopy alone. 
%
%
\end{abstract}
%
\keywords{
stars: formation -- ISM: kinematics and dynamics -- Herbig Ae/Be stars }
%
%
%
\section{Introduction} 
\label{sec:intro}
Herbig Ae/Be (HAeBe) stars \citep{herbig1960spectra} are intermediate-mass
pre-main-sequence stars ($\sim$2--8~$M_{\odot}$) surrounded by
circumstellar disks and, in many cases, winds and bipolar outflows or jets \citep{waters98,mottram2007difference,brittain23}.
Their inner circumstellar environments can comprise rotating disks, disk
winds, stellar winds, and other asymmetric structures, resulting in complex and evolving circumstellar geometries \citep{waters98}. These regions are generally too compact to be spatially resolved, making
indirect diagnostics of their geometry particularly important
\citep{vink2002probing,ababakr17}. In particular, $H\alpha$ emission
provides an important probe of the kinematics of circumstellar gas
\citep{waters98,oudmaijer1999halpha,vink2002probing,vink2005probing}.
Broad and double-peaked $H\alpha$ profiles can provide evidence for
rotating and/or outflowing gas; however, the line profile primarily traces
the velocity distribution of the emitting material and provides limited
constraints on its spatial geometry and asymmetry.

Linear spectropolarimetry provides a complementary probe of the
circumstellar environment because scattering by circumstellar material
produces polarization that depends on the spatial distribution and
geometry of the scattering medium \citep{clarke10}. $H\alpha$ spectropolarimetry of HAeBe stars probes 
the inner circumstellar environment on scales of a few stellar radii, providing a sensitive 
diagnostic of departures from spherical symmetry associated with the disk and 
wind regions \citep[e.g.,][]{vink2002probing}. 
Changes in polarization across $H\alpha$ can further constrain the relative spatial distributions of the line-emitting and scattering regions
\citep{vink2002probing,vink2005probing,mottram2007difference,
harrington2008spectropolarimetric,ababakr2016linear,ababakr17}.
Moreover, the wavelength dependence of the Stokes $Q$ and $U$ parameters 
allows the polarization to be examined as a function of velocity across the line. 
Velocity-dependent excursions or trajectories in the
$Q$--$U$ plane can therefore reveal asymmetric and
kinematically structured circumstellar material that may not be apparent 
from the $H\alpha$ intensity profile alone \citep[e.g.,][]{vink2002probing,ababakr17}.

Despite the extensive use of $H\alpha$ spectropolarimetry to probe asymmetric circumstellar structures in Herbig Ae/Be (HAeBe) stars, the temporal relationship between $H\alpha$ line-profile variability and changes in the associated polarization geometry remains poorly constrained. In particular, it is unclear whether changes in the kinematics of the line-emitting gas are accompanied by corresponding changes in the scattering geometry, or whether the two can exhibit an \emph{apparent kinematic--geometric decoupling} and evolve differently.

Here, we investigate this connection using multi-epoch $H\alpha$ spectropolarimetric observations obtained over $\sim$1--1.3~yr for three HAeBe stars: HD~58647, MWC~147, and MWC~120. These sources were first observed during the performance-verification phase of the in-house-developed medium-resolution spectropolarimeter ProtoPol \citep{srivastava2026development,maiti2026development} on the PRL 2.5-m telescope at Mt.~Abu, India, and were re-observed after about a year to investigate temporal variability. The sample was presented by \citet{maiti26}. Our inspection of the multi-epoch observations indicates differences in the temporal behavior of the $H\alpha$ line profiles and polarization signatures, motivating an investigation of possible \emph{kinematic--geometric decoupling} through the velocity-resolved $Q$--$U$ diagrams. Details of these sources are given in Section~\ref{sec:indvicom}. We compare the $H\alpha$ intensity profiles with the wavelength-dependent polarization degree, position angle, and $Q$--$U$ trajectories across epochs to examine whether changes in the line-emitting velocity structure are accompanied by corresponding changes in the polarization geometry.

The structure of this paper is as follows. Section~\ref{sec:obser}
presents the observational data sets used in this study. The observational
results are presented in Section~\ref{sec:xxdata}. In Section~\ref{sec:disc},
we discuss the implications of the derived results for the three selected
sources. Finally, Section~\ref{sec:sumcxf} summarizes our main conclusions.
\section{Observations and data reduction}
\label{sec:obser}
\subsection{ProtoPol and data reduction}
\label{sec:obser1a}
ProtoPol is a medium-resolution spectropolarimeter with a spectral resolution of 0.4--0.75~\AA{}, developed for the PRL 1.2-m and 2.5-m telescopes at Mt.~Abu, India \citep[][]{kumar2022designs, srivastava2024development, srivastava2026development, maiti2026development}. It is designed to cover the entire visible wavelength range from 4000 to 9600~\AA{}. ProtoPol employs a polarimeter unit, consisting of a half-wave plate (HWP) and a Wollaston prism (WP), coupled to a spectrometer unit. The spectrometer uses an echelle grating and a cross-disperser (CD) grating to produce cross-dispersed spectra of the ordinary and extraordinary (hereafter, o- and e-) rays on a $1\mathrm{K}\times1\mathrm{K}$ CCD across multiple echelle orders. Two interchangeable CD gratings, covering the blue (4000--6000~\AA{}) and red (6000--9600~\AA{}) wavelength ranges, allow observations across the full visible wavelength range. The instrument is also equipped with an integrated calibration unit containing a Uranium--Argon (UAr) lamp for wavelength calibration and a halogen lamp for tracing the echelle orders. ProtoPol has been extensively characterized through on-sky observations, with the instrumental polarization determined to be $\sim$0.1\%. Detailed results of the on-sky characterization are presented in \citet{maiti2026development}.

A complete spectropolarimetric data set consists of observations obtained at four HWP positions, namely $0^\circ$, $22.5^\circ$, $45^\circ$, and $67.5^\circ$, with calibration frames acquired after each science exposure. The raw ProtoPol data are reduced to science-ready products using an in-house, fully automated echelle spectropolarimetric data-reduction pipeline developed in Python using open-source libraries, including NumPy, SciPy, and Astropy. The reduction procedure includes bias and dark-frame subtraction, determination and tracing of the echelle orders, sky and scattered-light subtraction, extraction of the o- and e-ray intensities, wavelength calibration, and determination of the Stokes parameters from the intensities of the orthogonally polarized rays. Detailed descriptions of the ProtoPol data-reduction pipeline and its validation are presented in \citet{maiti2026development}.
\subsection{Target stars and previous spectropolarimetric observations}
\label{sec:indvicom}
\textbf{HD 58647:}
HD~58647 is a late-B-type Herbig star. Based on the variability of the $H\alpha$ centroid and equivalent width, \citet{baines2006binarity} suggested that the star may be a binary system. Multi-epoch spectropolarimetric observations by \citet{harrington2009spectropolarimetric} showed that its $H\alpha$ profile remained nearly invariant over the observed timescale. Polarimetric observations at multiple wavelengths led \citet{oudmaijer2001export} to suggest a disc-like circumstellar structure. Spectropolarimetric observations by \citet{vink2002probing} revealed an increase of $\sim0.6\%$ in polarization across the central absorption feature, while \citet{harrington2009spectropolarimetric} reported a similar increase of $\sim0.5\%$. The polarization across the blue- and redshifted emission peaks was found to be similar to the continuum polarization \citep{vink2002probing,harrington2009spectropolarimetric}.

\textbf{MWC 147:}
MWC~147 is classified as a B-type star. \citet{oudmaijer1999halpha} reported a continuum polarization of $\sim1.06\%$, together with a small depolarization of $\sim0.2\%$ relative to the continuum. Spectropolarimetric observations by \citet{vink2002probing} showed a decrease in polarization across the blue emission peak, although this effect was not explicitly classified as depolarization. Further spectropolarimetric observations were obtained by \citet{mottram2007difference}, albeit at lower spectral resolution. Multi-epoch observations with HiVIS and ESPaDOnS by \citet{harrington2009spectropolarimetric} showed little ($\sim0.2\%$) or no significant polarization difference between the blue- and redshifted portions of the $H\alpha$ profile.

\textbf{MWC 120:}
\citet{oudmaijer1999halpha} reported strong variability in both the $H\alpha$ intensity and polarization profiles of MWC~120, including changes in the relative strengths of the blue and red emission peaks and a clear change in polarization angle across the central absorption feature. An increase in polarization of $\sim0.4\%$ across the red emission peak was reported by \citet{vink2002probing}, whereas \citet{mottram2007difference} reported a smaller increase of $\sim0.2\%$. Subsequent multi-epoch spectropolarimetric observations with HiVIS and ESPaDOnS revealed substantial polarization variability, with changes of $\sim0.5$--$1.5\%$ in the Stokes $Q$ and $U$ spectra across the central absorption feature \citep{harrington2009spectropolarimetric}.
\subsection{Observations and data analysis}
\label{sec:obser1c}
The first-epoch observations of the two-epoch spectropolarimetric data presented here were obtained during November--December 2024 as part of the performance-verification program of ProtoPol. A second epoch was obtained during December 2025--March 2026 to investigate the temporal variability of the $H\alpha$ spectropolarimetric properties of the sources. The observational data of these Herbig Ae/Be stars were previously presented by \citet{maiti26}, who investigated their temporal variability. The data presented in this manuscript were adaptively binned to achieve polarization uncertainties of 0.1--0.3\%, thereby providing approximately uniform polarization uncertainties across the wavelength regions of interest. This binning is necessary because the signal strength varies substantially across the $H\alpha$ profile, particularly between emission peaks and absorption troughs. The uncertainties in the degree and position angle of polarization are given as $\sigma_P = \frac{1}{\sqrt{N/2}(S/N)}$ and $\sigma_{\theta} = \sigma_P/2p$ \citep[][]{patat2006error}, where N = 4 (for ProtoPol) is the number of HWP positions used to derive Stokes parameters, and (S/N) is the signal-to-noise ratio (SNR). The varying intensity across the $H\alpha$ profile consequently produces wavelength-dependent polarization uncertainties. We therefore employ an adaptive binning scheme in which the spectral bin width is increased dynamically until the required S/N, and hence the target $\sigma_P$, is achieved. This procedure ensures approximately constant polarization uncertainties across the binned spectra, at the expense of some degradation in spectral resolution.

The data have not been corrected for interstellar polarization (ISP), since its wavelength dependence is generally smooth across the relatively narrow $H\alpha$ region \citep{serkowski1974many}. Therefore, ISP can be treated as approximately constant across the line. Similarly, no correction for instrumental polarization has been applied. Extensive on-sky characterization of ProtoPol indicates an instrumental polarization level of $\sim0.1\%$ \citep{maiti2026development}. Observations of standard unpolarized and polarized stars over an 18-month period (December 2023--May 2025), encompassing the first-epoch observations, confirmed the temporal stability of the continuum polarization measurements and established typical uncertainties of $\sigma_Q$ and $\sigma_U \sim 0.5$--$0.6\%$ \citep{maiti2026development}. Similar observations of polarized standard stars were also conducted during the second-epoch observations, and their measurements were found to be consistent within the established error limits. Thus, both ISP and instrumental polarization are expected to introduce approximately constant polarization offsets to the data across the $H\alpha$ region. Consequently, any epoch-dependent changes in the continuum polarization near the $H\alpha$ emission are likely to be intrinsic to the sources.

We compare the continuum polarization measurements outside the $H\alpha$ line between the two epochs for each source to assess the temporal consistency of the observed polarization properties. Although the instrumental polarization has been established to be stable within the measurement uncertainties, the continuum flux is considerably lower than the $H\alpha$ line flux. Consequently, relatively broad wavelength bins were required in the continuum to achieve polarization uncertainties comparable to those in the line-emission region.

%
%
%
\section{Results}
\label{sec:xxdata}
%
We present the multi-epoch $H\alpha$ spectropolarimetric properties of HD~58647, MWC~120, and MWC~147. We obtained two-epoch spectropolarimetric observations of HD~58647 in December 2024 and March 2026, MWC~120 in November 2024 and December 2025, and MWC~147 in December 2024 and February 2026. The observations provide comparable temporal baselines of $\sim$1.0--1.3~yr for assessing changes in the inner circumstellar polarization signatures of these sources \citep{maiti26}. For each source, we examine the $H\alpha$ intensity profiles, wavelength- or velocity-dependent polarization degree ($P$) and position angle (PA or $\theta$), and the corresponding velocity-resolved $Q$--$U$ diagrams.

\begin{figure*}
\center
\includegraphics[width=\textwidth]{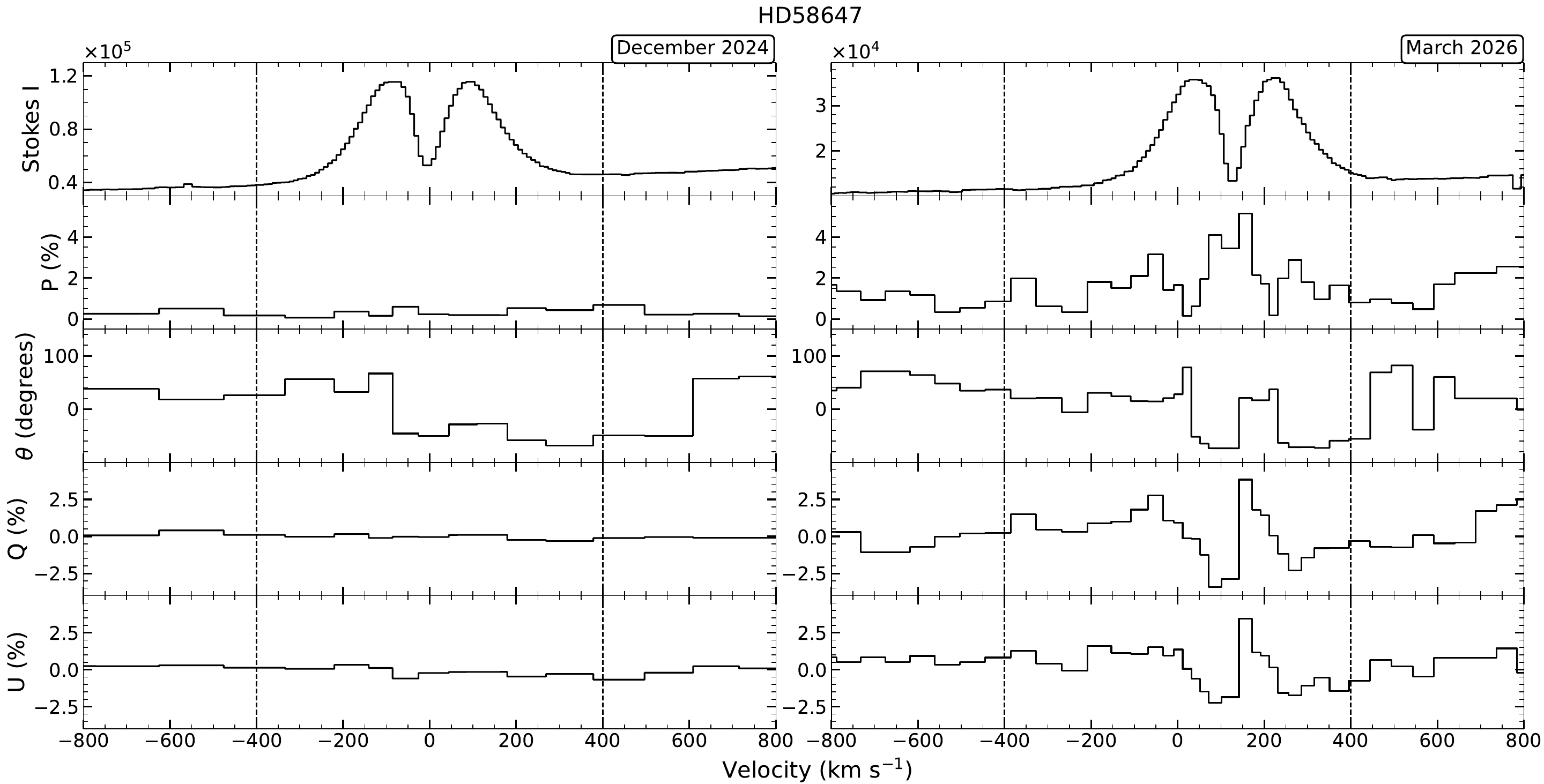}
\caption{Multi-epoch $H\alpha$ spectropolarimetry of the Herbig Ae/Be star, HD~58647. 
The two epochs are shown in the left and right columns, respectively. For each epoch, the
panels (from top to bottom) display the normalized $H\alpha$ intensity, degree of linear
polarization ($P$), polarization position angle ($\theta$), and the Stokes
parameters $Q$ and $U$ as a function of velocity. The vertical dashed lines
indicate the adopted $H\alpha$ velocity range. The flux and polarization
spectra are binned using variable bin sizes to achieve approximately
constant polarization uncertainties. 
The resulting polarization uncertainties are 0.1\% and 0.3\% for the first and second epochs, respectively. The plots of $P$, $\theta$, $Q$, and $U$ are presented using the same axis ranges within each epoch. However, the Stokes $I$ profiles are displayed over different value ranges between the two epochs.}
\label{nufg1}
\end{figure*}

\begin{figure*}
\center
\includegraphics[width=\textwidth]{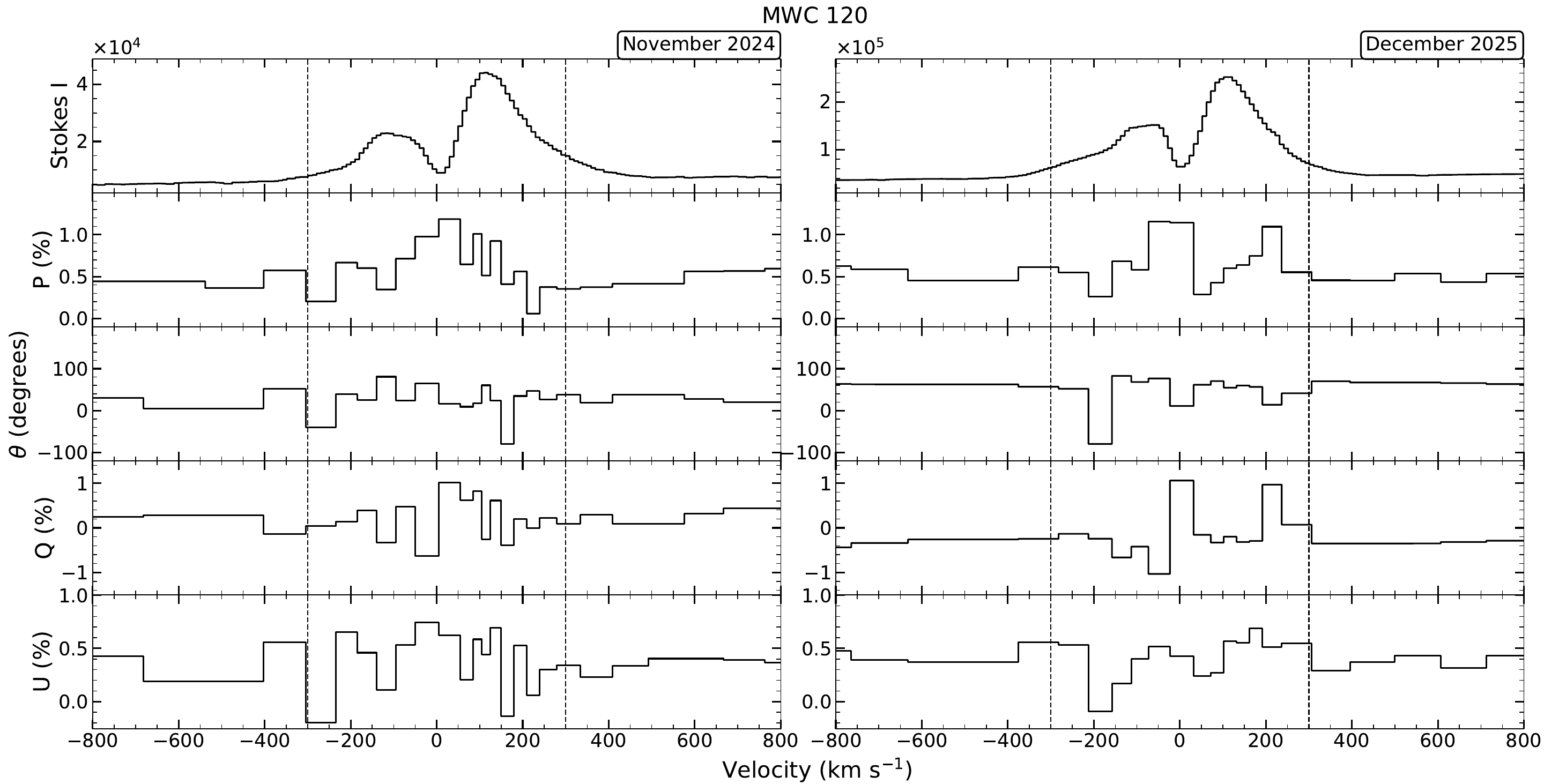}
\caption{Same as Figure~\ref{nufg1}, but for MWC~120. The two epochs are shown in the left and right columns, respectively. The resulting polarization uncertainties for the first and second epochs are 0.2\% and 0.1\%, respectively. The plots of $P$, $\theta$, $Q$, and $U$ are shown using the same axis ranges within the two epochs. But, the Stokes $I$ profiles are presented over different value ranges between the two epochs.}
\label{nufg2}
\end{figure*}

\begin{figure*}
\center
\includegraphics[width=\textwidth]{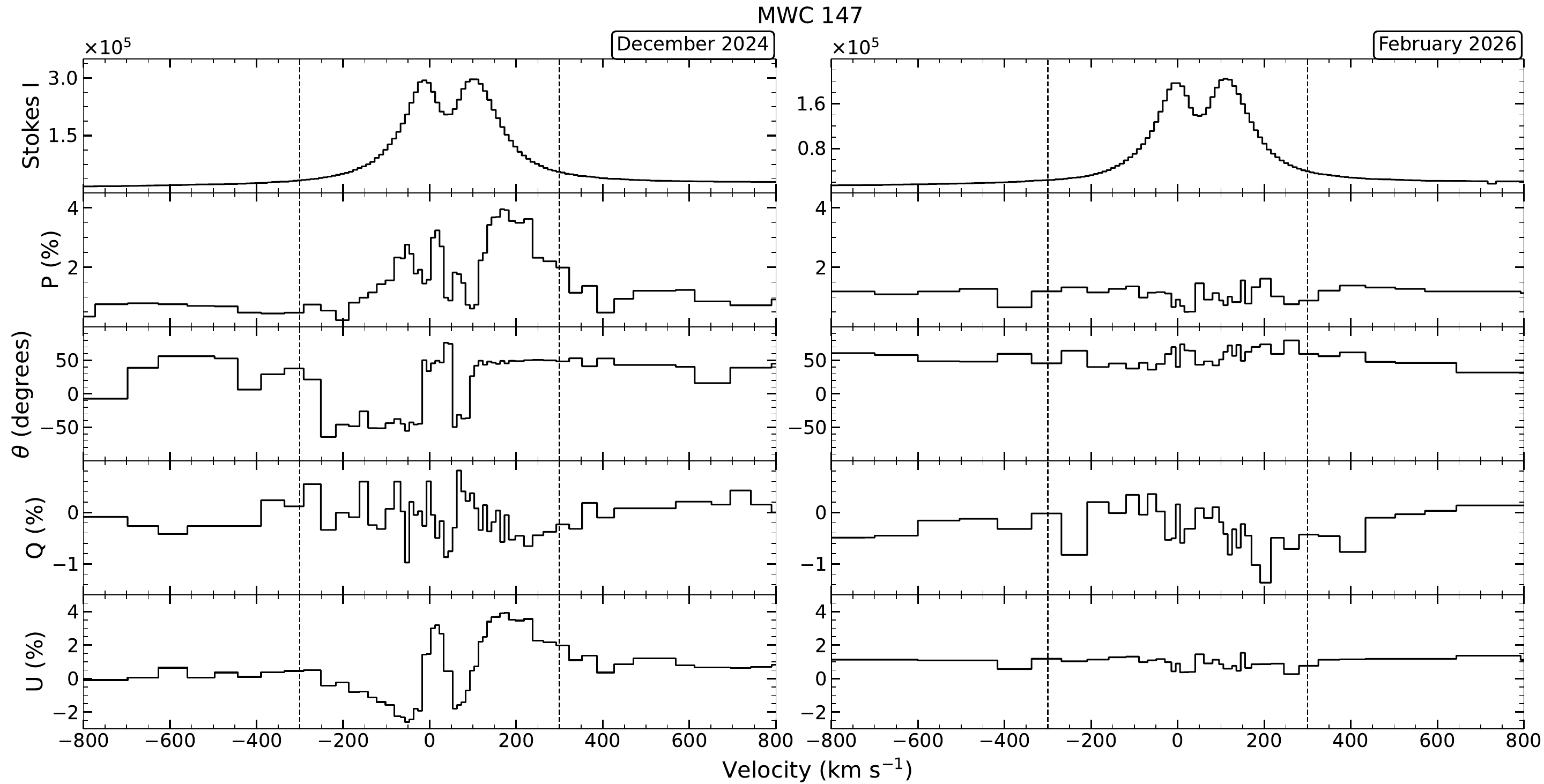}
\caption{Same as Figure~\ref{nufg1}, but for MWC~147. The two epochs are presented in the left and right columns, respectively. The resulting polarization uncertainties for the first and second epochs are 0.2\% and 0.2\%, respectively. The plots of $P$, $\theta$, $Q$, and $U$ are presented using the same axis ranges within each epoch. However, the Stokes $I$ profiles are not displayed over identical value ranges between the two epochs.}
\label{nufg2a}
\end{figure*}

Figures~\ref{nufg1},~\ref{xnufg1},~\ref{nufg2},~\ref{nufg2a},~\ref{xnufg2}, and~\ref{xnufg2a} present the spectropolarimetric results for the three sources. The $H\alpha$ intensity profiles and corresponding polarization spectra are shown in Figures~\ref{nufg1},~\ref{nufg2}, and~\ref{nufg2a}, while the velocity-resolved $Q$--$U$ diagrams are presented in Figures~\ref{xnufg1},~\ref{xnufg2}, and~\ref{xnufg2a}. For each source, the $P$, $\theta$, $Q$, and $U$ plots use identical axis ranges within each epoch, whereas the Stokes $I$ profiles are displayed over different value ranges between the two epochs (see Figures~\ref{nufg1},~\ref{nufg2}, and~\ref{nufg2a}).
%
%
In the $Q$--$U$ plane, a trajectory traces the path followed by the velocity-ordered Stokes $Q$ and $U$ measurements across the $H\alpha$ line profile, with each point representing a different velocity interval. Thus, the trajectory characterizes the velocity-dependent variation in the observed polarization. Its extent represents the amplitude of the excursion in the $Q$--$U$ plane rather than the physical size of the circumstellar scattering region. The morphology and orientation of these trajectories provide observational constraints on the projected polarization structure, although they do not uniquely determine the underlying scattering geometry.

In Figures~\ref{nufg1},~\ref{nufg2}, and~\ref{nufg2a}, the vertical dashed lines indicate the adopted $H\alpha$ velocity range. The continuum polarization measured outside the $H\alpha$ line velocity interval provides a reference for comparing the two epochs. The continuum polarization degree and PA are consistent within the measurement uncertainties, suggesting that the observed changes in the velocity-resolved polarization are not accompanied by comparable changes in the continuum polarization. 
%
\subsection{HD~58647}
\label{sec:hd58647}
\begin{figure*}
\center
\includegraphics[width=\textwidth]{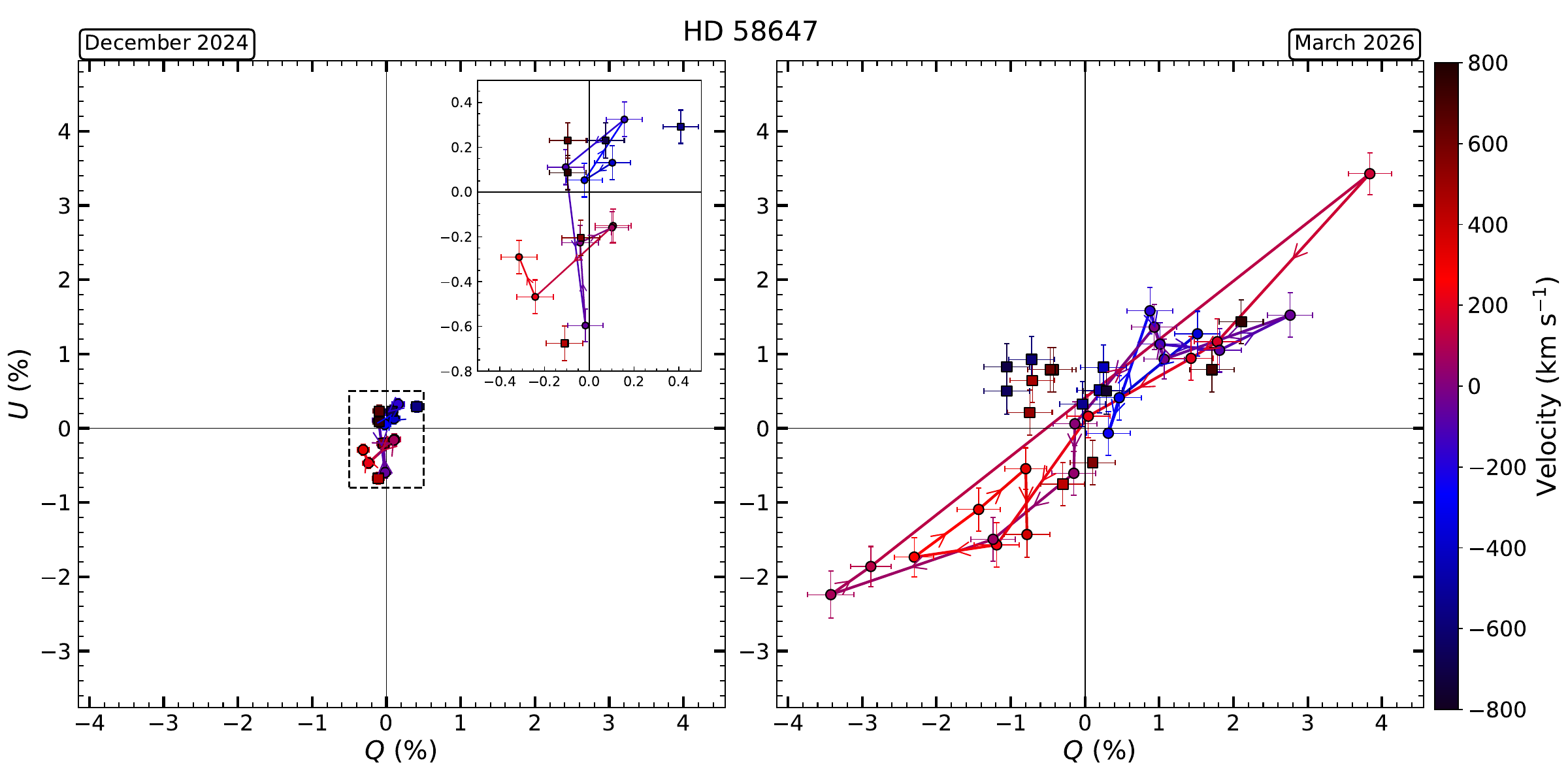}
\caption{Velocity-resolved $Q$--$U$ diagrams of the Herbig Ae/Be star HD~58647. The diagrams for the two epochs are presented in the left and right columns, respectively. In the left panel, the top-right inset shows a zoomed-in view of the region enclosed by the dashed box. In all panels, including the inset, the colored squares denote the continuum polarization, while the colored circles and connecting lines trace the polarization across the $H\alpha$ line profile as a function of line-of-sight velocity. Each circle represents a polarization measurement averaged over a velocity bin, with variable bin widths adopted to achieve approximately constant polarization uncertainties. The color scale on the right indicates the corresponding velocity (see also Figure~\ref{nufg1}). Error bars represent the uncertainties in the binned polarization measurements; where not readily visible, they are comparable to or smaller than the symbol size. The black horizontal and vertical lines indicate $Q=0$ and $U=0$, respectively. The typical polarization uncertainties are 0.1\% and 0.3\% for the first and second epochs, respectively.}
\label{xnufg1}
\end{figure*}

\begin{figure*}
\center
\includegraphics[width=\textwidth]{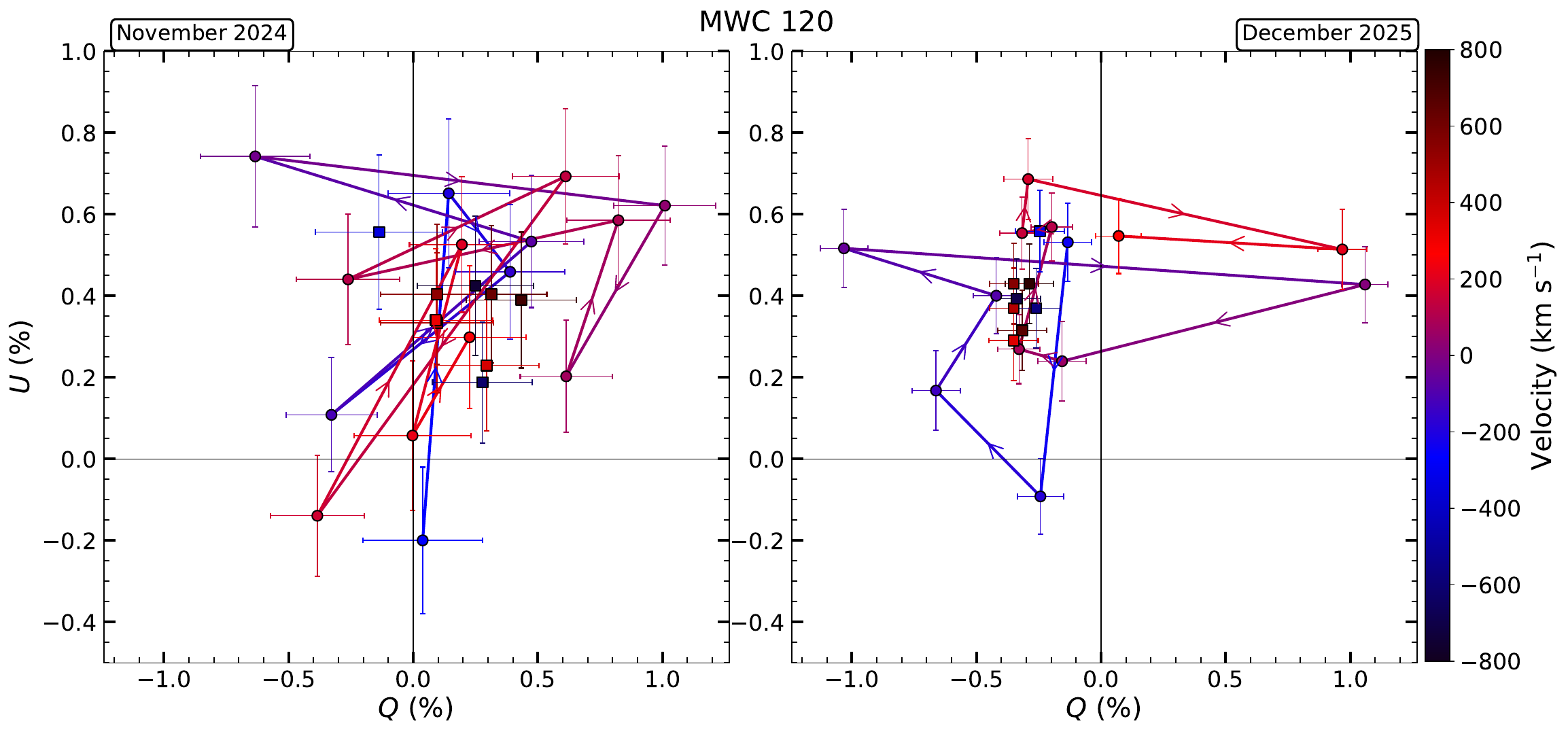}
\caption{Same as Figure~\ref{xnufg1}, but for MWC~120. 
The diagrams for the two epochs are presented in the left and right columns, respectively (see also Figure~\ref{nufg2}). The resulting polarization uncertainties for the first and second epochs are 0.2\% and 0.1\%, respectively.}
\label{xnufg2}
\end{figure*}

\begin{figure*}
\center
\includegraphics[width=\textwidth]{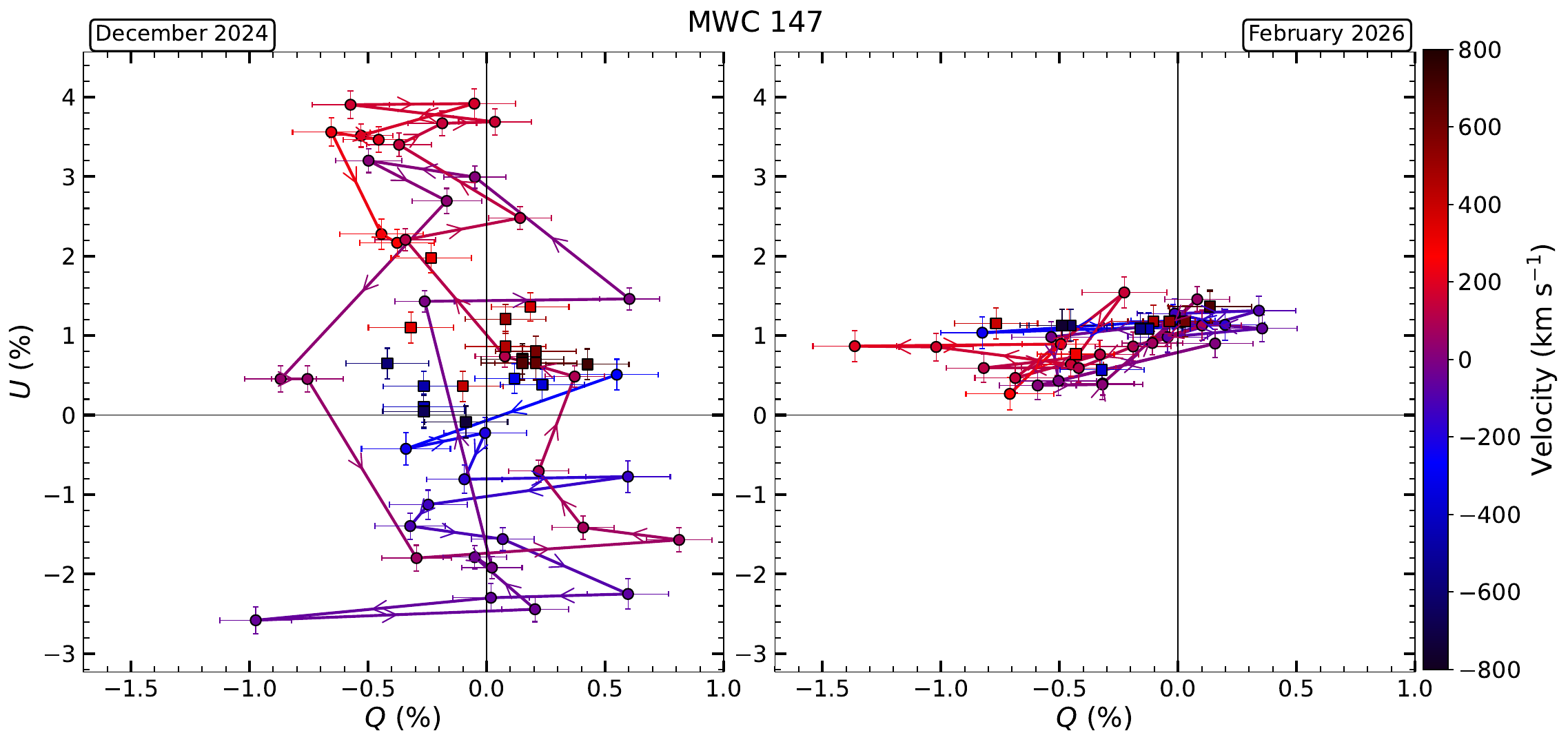}
\caption{Same as Figure~\ref{xnufg1}, but for MWC~147. 
The diagrams for the two epochs are presented in the left and right columns, respectively (see also Figure~\ref{nufg2a}). The resulting polarization uncertainties for the first and second epochs are 0.2\% and 0.2\%, respectively.}
\label{xnufg2a}
\end{figure*}

The multi-epoch $H\alpha$ spectropolarimetric observations of HD~58647, obtained in December 2024 and March 2026, are presented in Figures~\ref{nufg1} and~\ref{xnufg1}. The $H\alpha$ intensity profiles exhibit broadly comparable velocity extents at the two epochs, although noticeable differences in line strength and peak intensity are apparent (see Figure~\ref{nufg1}).

The polarization spectra exhibit epoch-dependent variations across the $H\alpha$ line profile, with velocity-dependent changes in both polarization degree and PA (see Figure~\ref{nufg1}). Similar variations are evident in the corresponding $Q$ and $U$ spectra. A near-$90^\circ$ change in PA is observed between the blue- and red-shifted portions of the line, indicating a substantial change in the net polarization direction across the profile. Such behavior may arise from variations in the relative contributions of different polarization components, including continuum and line polarization, which combine vectorially in the Stokes $Q$--$U$ plane. Near-$90^\circ$ PA flips have also been associated with mirror-symmetric circumstellar scattering geometries \citep{schmid1998raman}. However, the observed PA change alone does not establish the presence of multiple spatially distinct scattering regions or uniquely constrain the system inclination and three-dimensional scattering geometry. In the March 2026 observations, multiple near-$90^\circ$ changes in PA are apparent across the line profile, indicating more complex velocity-dependent polarization behavior. These changes may reflect epoch-dependent variations in the relative contributions of different polarizing components within the inner circumstellar environment \citep{vink2005polarimetric}.

These variations are also reflected in the corresponding $Q$--$U$ diagrams (see Figure~\ref{xnufg1}). The December 2024 observations show a relatively compact trajectory, with the velocity-resolved measurements concentrated within a limited region of the $Q$--$U$ plane. The enlarged view of the December 2024 $Q$--$U$ diagram (top-right inset in the left panel of Figure~\ref{xnufg1}) reveals distinct velocity-dependent excursions of the polarization points around the continuum polarization position. In particular, the negative- and positive-velocity portions of the line profile occupy different regions of the $Q$--$U$ plane, producing a complex, non-linear trajectory. In contrast, the March 2026 observations exhibit a substantially more extended, approximately diagonal trajectory, spanning a much larger range in both Stokes parameters. This difference highlights a substantial change in the velocity-resolved polarization morphology between the two epochs. The pronounced velocity dependence further indicates that different portions of the $H\alpha$ profile are associated with different polarization degrees and position angles. This behavior may reflect a structured spatial relationship between the line-emitting and scattering regions, although the underlying circumstellar geometry cannot be uniquely constrained from the observed $Q$--$U$ trajectories alone.
\subsection{MWC~120}
\label{sec:mwc120}
The multi-epoch $H\alpha$ spectropolarimetric observations of MWC~120, obtained in November 2024 and December 2025, are presented in Figures~\ref{nufg2} and~\ref{xnufg2}. The $H\alpha$ intensity profiles retain broadly similar double-peaked morphologies and overall velocity extents at the two epochs, although variations in the relative strengths of the line components and peak intensities are apparent (see Figure~\ref{nufg2}).

The polarization spectra exhibit velocity-dependent variations in both polarization degree and PA at the two epochs. The corresponding $Q$--$U$ diagrams show structured and open trajectories in both observations (see Figure~\ref{xnufg2}). In November 2024, the trajectory displays a relatively compact, multidirectional distribution of velocity-resolved measurements, with excursions along both Stokes parameters. In December 2025, the trajectory remains open but exhibits a different distribution, characterized by a more pronounced horizontal elongation in the $Q$--$U$ plane.

Unlike HD~58647, where the most evident change is the development of a pronounced linear excursion, MWC~120 exhibits changes primarily in the orientation and distribution of its open trajectory between the two epochs. Nevertheless, velocity-dependent variations in polarization remain evident in both observations. These changes indicate temporal variability in the projected polarization structure, despite the broadly comparable overall velocity extent of the $H\alpha$ emission.
\subsection{MWC~147}
\label{sec:mwc147}
The multi-epoch $H\alpha$ spectropolarimetric observations of MWC~147, obtained in December 2024 and February 2026, are presented in Figures~\ref{nufg2a} and~\ref{xnufg2a}. The $H\alpha$ intensity profiles retain broadly similar double-peaked morphologies and overall velocity extents at the two epochs, although differences in the detailed line structure and relative strengths of the two components are apparent (see Figure~\ref{nufg2a}).

The polarization spectra exhibit pronounced velocity-dependent variations, with epoch-dependent differences in both polarization degree and PA. A near-$90^\circ$ change in PA is observed across the line during the first epoch, including a rotation extending redward of the $H\alpha$ emission. This behavior may be indicative of extended circumstellar scattering regions and possible mirror symmetries in the system \citep{schmid1998raman}. However, since ISP and instrumental polarization are expected to remain approximately constant across this spectral region, they cannot independently account for the wavelength-dependent PA rotation. The observed behavior may instead reflect changes in the net polarization resulting from the wavelength-dependent vector combination of intrinsic continuum and line polarization components. The corresponding $Q$ and $U$ spectra also display markedly different velocity-dependent structures between the epochs. These changes are reflected in the $Q$--$U$ diagrams, which reveal a substantial evolution in trajectory morphology (see Figure~\ref{xnufg2a}). The December 2024 observations show a complex and vertically extended trajectory, with substantial excursions, particularly along the $U$ direction. In February 2026, the trajectory becomes considerably more confined and exhibits a comparatively coherent, approximately linear arrangement with a predominantly horizontal orientation in the $Q$--$U$ plane.

Thus, MWC~147 exhibits a marked change in its velocity-resolved polarization morphology, from a complex, vertically extended trajectory to a more compact and predominantly horizontally elongated one. The more confined trajectory in February 2026 indicates a reduced amplitude and less complex distribution of the observed variations in polarization in the $Q$--$U$ plane. These changes occur despite the broadly comparable overall velocity extent of the $H\alpha$ emission at the two epochs.

\begin{table*}
\centering
\large
\caption{Observed spectropolarimetric morphologies of the three Herbig Ae/Be stars (HD~58647, MWC~120, and MWC~147; see Figures~\ref{nufg1}--\ref{xnufg2a} and also Section~\ref{sec:xxdata}). The table summarizes the epoch-dependent characteristics of the velocity-resolved $Q$--$U$ trajectories, their possible interpretations in terms of the projected polarization structure, and the observed temporal changes. The geometrical interpretations are not unique.}
\label{tab:scattering_geometry}

\resizebox{\textwidth}{!}{%
\begin{tabular}{lccc}
\hline\hline
Source &
\shortstack{Epoch-dependent $Q$--$U$\\trajectory morphology} &
\shortstack{Possible interpretation of projected\\polarization structure} &
\shortstack{Observed temporal behavior} \\
\hline

HD~58647 &
\shortstack[l]{December 2024: Compact distribution near the origin,\\without a clearly defined linear excursion.\\
March 2026: Pronounced, approximately linear and\\velocity-ordered trajectory.} &
\shortstack[l]{December 2024: Relatively small velocity-dependent\\variations in polarization.\\
March 2026: More pronounced and organized variations\\in polarization degree and position angle.} &
\shortstack[l]{A pronounced linear excursion emerges in March 2026,\\indicating a substantial change in the velocity-dependent\\polarization morphology.} \\

\hline

MWC~120 &
\shortstack[l]{November 2024: Relatively compact, structured,\\multidirectional trajectory with excursions in both $Q$ and $U$.\\
December 2025: Open, more horizontally elongated\\trajectory with a different distribution.} &
\shortstack[l]{November 2024: Structured velocity-dependent\\variations in polarization.\\
December 2025: Preferential orientation of variations\\along the $Q$ direction.} &
\shortstack[l]{The trajectory remains open at both epochs, but its\\orientation and distribution change, while velocity-dependent\\variations in polarization persist.} \\

\hline

MWC~147 &
\shortstack[l]{December 2024: Complex, vertically extended trajectory\\with large excursions, particularly in $U$.\\
February 2026: More confined, comparatively coherent,\\approximately linear trajectory with a horizontal orientation.} &
\shortstack[l]{December 2024: Strong velocity-dependent variations\\with substantial excursions in the $Q$--$U$ plane.\\
February 2026: More confined variations in projected\\polarization, with a reduced amplitude.} &
\shortstack[l]{The trajectory undergoes substantial contraction and\\simplification, indicating a reduction in the amplitude and\\complexity of the velocity-dependent variations.} \\

\hline\hline
\end{tabular}%
}
\end{table*}
%
%
\section{Discussion}
\label{sec:disc}
The $H\alpha$ line is a useful diagnostic of the inner circumstellar environments of Herbig Ae/Be stars. Linear spectropolarimetry provides indirect spatial information because stellar continuum photons can undergo Thomson scattering by free electrons in ionized circumstellar material close to the star \citep{vink2002probing,ababakr17}. Since the observed polarization depends on the projected distribution of the scattering material, changes in polarization across an emission line can reveal spatially and kinematically dependent structure that is not apparent in the total-intensity spectrum alone. In particular, excursions and structured trajectories in the $Q$--$U$ plane indicate variations in the observed polarization properties across different velocity intervals of the line \citep{vink2002probing,ababakr17}. Such signatures have been interpreted in terms of flattened or rotating circumstellar structures, although a given $Q$--$U$ morphology does not uniquely determine the underlying physical geometry \citep{vink2002probing}. Furthermore, the observed polarization results from the vector combination of continuum and line polarization components. Variations in their relative contributions can therefore produce substantial changes in polarization degree and position angle without necessarily requiring multiple spatially distinct scattering regions.

Previous spectropolarimetric studies of Herbig Ae/Be stars have established that polarization changes across $H\alpha$ are common and provide evidence for departures from spherical symmetry in their inner circumstellar environments \citep{vink2002probing,vink2005probing,mottram2007difference,ababakr2016linear,ababakr17}. However, much of this work has provided individual-epoch or limited multi-epoch views of the relationship between the line-emitting and scattering regions. The present observations extend these investigations by examining temporal changes in the velocity-resolved polarization signatures of the same three stars over baselines of $\sim$1--1.3~yr.

A comparison of the intensity and polarization spectra reveals a notable difference in their temporal behavior. Although the $H\alpha$ profiles retain broadly comparable overall morphologies and velocity extents between the two epochs in all three sources, variations in line strength and peak intensity are apparent. In contrast, the velocity-resolved polarization signatures exhibit distinct source-dependent changes (see Figures~\ref{xnufg1},~\ref{xnufg2},~\ref{xnufg2a}, and also Table~\ref{tab:scattering_geometry}). HD~58647 shows a transition from a compact, weakly structured $Q$--$U$ trajectory to a pronounced, predominantly linear excursion. MWC~120 retains open and structured trajectories at both epochs, but their orientation and distribution change. MWC~147 exhibits a transition from a complex, vertically extended trajectory to a more compact and comparatively coherent, predominantly horizontally elongated one (see Table~\ref{tab:scattering_geometry} and also Section~\ref{sec:xxdata}). These differences demonstrate that substantial changes in the observed polarization structure can occur without corresponding changes in the broad morphology or overall velocity extent of the $H\alpha$ emission. We refer to this observational contrast as an \emph{apparent spectroscopic--polarimetric decoupling}. This terminology does not imply physical independence between the line-emitting and scattering regions or establish a unique change in their three-dimensional geometry.

The observed diversity may reflect temporal variations in the projected scattering environment and/or changes in the relative contributions of different polarizing components within the inner circumstellar system. The stellar vicinity, disc, disc atmosphere, and disc wind may all contribute to the observed polarization. Changes in density, optical depth, illumination, or local asymmetries could modify the velocity-resolved polarization signatures without producing comparably large changes in the overall $H\alpha$ line-profile morphology. However, the present unresolved observations do not allow us to distinguish uniquely between changes in disc structure, disc-wind contributions, localized density inhomogeneities, or other variations in the inner circumstellar material.

The main observational result of this study is that the velocity-resolved polarization properties of Herbig Ae/Be stars can evolve substantially between epochs even when the broad morphology and overall velocity extent of their $H\alpha$ emission remain relatively stable. Multi-epoch, velocity-resolved spectropolarimetry therefore provides information complementary to the $H\alpha$ intensity profile, enabling temporal changes in the projected polarization structure of unresolved circumstellar environments to be investigated. A larger sample with denser temporal sampling and complementary high-angular-resolution observations will be required to identify the physical processes responsible for this evolution.
\section{Summary and Conclusions}
\label{sec:sumcxf}
We have presented multi-epoch spectropolarimetric observations of three
Herbig Ae/Be stars, HD~58647, MWC~120, and MWC~147, obtained over timescales of $\sim$1--1.3~yr. 
By combining $H\alpha$ line profiles,
polarization spectra, and velocity-resolved $Q$--$U$ diagrams, we
investigated temporal changes in their unresolved inner circumstellar
environments. Our main results are summarized below.

\begin{enumerate}

\item The $H\alpha$ profiles retain broadly similar overall morphologies
and velocity extents between the two epochs in all three sources,
although noticeable variations in line strength and peak intensity
are observed. Thus, the broad spectroscopic characteristics remain
relatively stable over the observed timescales.

\item The polarization properties show pronounced and source-dependent
temporal variability, involving changes in polarization degree,
position angle, and the morphology and orientation of the
velocity-resolved $Q$--$U$ trajectories. The three sources exhibit
distinct temporal behaviors: HD~58647 shows a transition from a
compact, weakly structured trajectory to a pronounced, predominantly
linear excursion; MWC~120 exhibits open, structured trajectories
at both epochs, with changes in their orientation and distribution;
and MWC~147 shows a transition from a complex, vertically extended
trajectory to a considerably more compact, predominantly horizontally
elongated trajectory.

\item The velocity-resolved polarization provides constraints on the
projected polarization structure of the inner circumstellar
environment. In HD~58647, the pronounced velocity dependence indicates
that different velocity intervals across the $H\alpha$ profile are
associated with distinct polarization degrees and position angles,
suggesting a structured spatial relationship between the line-emitting
and scattering regions. The epoch-dependent changes in the $Q$--$U$
trajectories further reveal temporal variations in the projected
polarization properties.

\item A key observational result is the substantial temporal change
in velocity-resolved polarization morphology despite broadly
comparable $H\alpha$ profile shapes and overall velocity extents.
We describe this contrast as an \emph{apparent
spectroscopic--polarimetric decoupling}. This terminology refers
to the difference between the relatively stable broad
$H\alpha$ profiles and the more strongly evolving velocity-resolved
polarization signatures. It does not imply physical independence
of the line-emitting and scattering regions or establish a unique
change in their three-dimensional geometry.

\end{enumerate}

Overall, the observations demonstrate that the velocity-resolved
polarization properties of the inner circumstellar environments of
these Herbig Ae/Be stars can change substantially between epochs,
even when the broad $H\alpha$ profile morphology and overall velocity
extent remain relatively stable. Multi-epoch, velocity-resolved
spectropolarimetry therefore provides a complementary probe of
temporal changes in the projected polarization structure of
unresolved circumstellar environments, revealing variations that
may not be readily apparent from spectroscopy alone. Although the
three-object sample does not uniquely establish the physical origin
of the observed variability, these results provide observational
constraints on the temporal evolution of the inner circumstellar
environments of Herbig Ae/Be stars.
\section*{Acknowledgments}
We sincerely thank the anonymous referee for constructive suggestions, which have led to significant improvements in this manuscript.  
The research work at the Physical Research Laboratory is funded by the Department of Space, Government of India. AM and SSG gratefully acknowledge PRL for their Ph.D. research fellowships. PRL operates the 1.2 m and 2.5 m telescopes at the Mt. Abu Observatory. We acknowledge the use of data obtained with the ProtoPol instrument from both the PRL 1.2 m and 2.5 m telescopes at the Mt. Abu Observatory.
%
%
\bibliography{reference}{}
\bibliographystyle{aasjournal}
\end{document}